\documentclass[sigconf]{acmart}

\AtBeginDocument{%
  }

\setcopyright{acmlicensed}
\copyrightyear{2018}
\acmYear{2018}
\acmDOI{XXXXXXX.XXXXXXX}

\acmConference[Conference acronym 'XX]{Make sure to enter the correct
  conference title from your rights confirmation emai}{June 03--05,
  2018}{Woodstock, NY}
\acmISBN{978-1-4503-XXXX-X/18/06}

\usepackage{graphicx}
\usepackage{multirow}
\usepackage{tikz}
\usepackage{makecell}
\usepackage{algorithm, algpseudocode}
\usepackage{amsmath}
\usepackage{array}
\usetikzlibrary{patterns}
\usepackage{subcaption}
\usepackage{multirow}
\usepackage{graphicx}
\usepackage[normalem]{ulem}
\useunder{\uline}{\ul}{}

\begin{document}
\nocite{*}

\title{Sparser Training for On-Device Recommendation Systems}

\author{Yunke Qu}
\affiliation{%
  \institution{The University of Queensland}
  \city{Brisbane}
  \state{}
  \country{Australia}
}
\email{yunke.qu@uq.net.au}

\author{Liang Qu}
\affiliation{
    \institution{Southern University of Science and Technology}
    \city{Shenzhen}
    \state{}
    \country{China}
}
\email{qul@mail.sustech.edu.cn}

\author{Tong Chen}
\affiliation{%
  \institution{The University of Queensland}
  \city{Brisbane}
  \state{}
  \country{Australia}
}
\email{tong.chen@uq.edu.au}

\author{Xiangyu Zhao}
\affiliation{
  \institution{City University of Hong Kong}
  \city{Hong Kong}
  \state{}
  \country{}
}
\email{xy.zhao@cityu.edu.hk}

\author{Jianxin Li}
\affiliation{
  \institution{Edith Cowan University}
  \city{Perth}
  \state{}
  \country{Australia}
}
\email{jianxin.li@ecu.edu.au}

\author{Hongzhi Yin}
\authornote{Corresponding author}
\affiliation{%
  \institution{The University of Queensland}
  \city{Brisbane}
  \state{}
  \country{Australia}
}
\email{h.yin1@uq.edu.au}

\renewcommand{\shortauthors}{Yunke Qu et al.}

\begin{abstract}
Recommender systems often rely on large embedding tables that map users and items to dense vectors of uniform size, leading to substantial memory consumption and inefficiencies. This is particularly problematic in memory-constrained environments like mobile and Web of Things (WoT) applications, where scalability and real-time performance are critical. Various research efforts have sought to address these issues. Although embedding pruning methods utilizing Dynamic Sparse Training (DST) stand out due to their low training and inference costs, consistent sparsity, and end-to-end differentiability, they face key challenges. Firstly, they typically initializes the mask matrix, which is used to prune redundant parameters, with random uniform sparse initialization. This strategy often results in suboptimal performance as it creates unstructured and inefficient connections. Secondly, they tend to favor the users/items sampled in the single batch immediately before weight exploration when they reactivate pruned parameters with large gradient magnitudes, which does not necessarily improve the overall performance. Thirdly, while they use sparse weights during forward passes, they still need to compute dense gradients during backward passes. In this paper, we propose SparseRec, an lightweight embedding method based on DST, to address these issues. Specifically, SparseRec initializes the mask matrix using Nonnegative Matrix Factorization. It accumulates gradients to identify the inactive parameters that can better improve the model performance after activation. Furthermore, it avoids dense gradients during backpropagation by sampling a subset of important vectors. Gradients are calculated only for parameters in this subset, thus maintaining sparsity during training in both forward and backward passes. We evaluate sparseRec on two public datasets, demonstrating performance improvements of up to 11.79\% across three base recommenders and three different density ratios.

\end{abstract}


\maketitle

\section{Introduction} \label{sec:intro}
Recommender systems (RSs) are widely applied in mobile and web applications. Typically, RSs are deployed on cloud servers and learn user/item embeddings based on user-item interactions. These embeddings are subsequently utilized to predict users' preferred items by computing similarity scores between user and item embeddings. However, as the number of users and items increases, this cloud-based architecture encounters significant limitations, including privacy risks, high resource consumption and increased latency \cite{Xia2023EOD, Xia2022ODRec, Wang2020NextPoint}.

In light of these limitations, there is an increasing demand for deploying recommender systems directly on devices, such as mobile phones and the Web of Things (WoT). This shift has led to a greater emphasis on lightweight recommendation solutions that offer low latency and enhanced privacy by processing data locally, known as on-device recommender systems (ODRSs), which primarily focus on the efficient deployment and updating of lightweight embedding tables on resource-constrained devices.

Research in this area can be broadly categorized into three main approaches: (1) parameter-sharing methods including compositional embeddings \cite{Liang2023CERP, Liang2024LEGCF, Kang2021DHE, Shi2020CompositionalEmb}, (2) variable-size embedding methods \cite{Liu2020ESAPN, Qu2023CIESS, Qu2024SCALL} and (3) embedding pruning methods \cite{Liu2021PEP, Wang2024DSL, Wang2022LTH}. Parameter-sharing methods enable different users and items to share embedding parameters, though they still rely on dense embeddings, limiting memory efficiency. Variable-size embedding methods assign different dimensions to users and items based on frequency and contextual information, often utilizing AutoML techniques \cite{Zheng2024AutoMLSurvery}, but they typically require long training times. In contrast, embedding pruning methods eliminate redundant parameters from the dense embedding table while maintaining performance. Embedding pruning offers two key advantages over parameter-sharing and variable-size embedding methods: first, forward passes operate on sparse embeddings, unlike parameter-sharing methods; second, embedding pruning is end-to-end differentiable and doesn't require iterative training for performance evaluation, unlike variable-size embedding approaches. Furthermore, embedding pruning methods (e.g., DSL \cite{Wang2024DSL}) based on the Dynamic Sparse Training (DST) paradigm provide the benefits of maintaining consistent sparsity throughout training, establishing them as the favored state-of-the-art approach for lightweight recommendation.

Despite the improvements in training time and memory efficiency offered by recent embedding pruning methods, they still face three key drawbacks due to the limitations of the DST paradigm they adopt: (1) poor sparse initialization, (2) an ineffective regrowth strategy and (3) the use of dense gradients. First, traditional DST initializes the mask matrix, which is used to prune parameters, with random uniform initialization. This strategy removes a certain number of embedding parameters to conform to a predefined density ratio. Each parameter has an equal chance of being pruned. Various studies \cite{Peng2022SparseGraph, Evci2020RigL} have shown that random uniform sparse initialization leads to suboptimal post-training performance because it creates unstructured and inefficient connections. Alternatives like Erdös-Rényi (ER) initialization, which generates layer-wise sparse masks in DNNs and CNNs \cite{Mocanu2018Scalable, Evci2020RigL}, are not applicable to GNNs-based recommenders, where the pruning focuses on the embedding table rather than dense or convolution layers. On the other hand, gradient-based methods employ dense gradients as importance scores to prune parameters \cite{Lee2018Snip, Wang2020GraSp, Tanaka2020SynFlow}, increasing computational overhead and negating some of the efficiency benefits of DST. Second, traditional DST adopts the conventional regrowth strategy of reinstating a portion of parameters with the largest instantaneous gradient magnitudes. While this strategy is effective in CNNs and DNNs, it leads to suboptimal performance in recommendation settings because embedding vectors for users/items that are sampled in the current batch tend to have large magnitudes. As a result, the parameters reactivated by traditional DST primarily belong to the sampled users/items in that batch, which are not necessarily important for overall performance improvement. Third, traditional DST computes dense gradients in backward passes, incurring a dense gradient table during the training process.

To overcome the aforementioned challenges, we propose SparseRec, an lightweight embedding method that advances state-of-the-art embedding pruning methods by addressing the limitations of the DST paradigm. Firstly, SparseRec initializes the mask matrix with Nonnegative Matrix Factorization (NMF) to achieve sparse initialization. This output mask matrix of NMF is naturally sparse and, contrary to random uniform initialization, this data-driven method initializes the mask matrix to a local optimum, effectively addressing (1). Secondly, it accumulates gradients over several consecutive batches to regrow inactive embedding parameters. Compared to instantaneous gradients from a single batch, cumulative gradients can better identify the inactive parameters that can significantly improve the model performance after activation, tackling (2). Thirdly, SparseRec utilizes sparse gradients, which directly addresses (3). To achieve sparse gradients, Heddes et al. \cite{Heddes2024AlwaysSparse} samples and reactivates inactive parameters likely to have large gradients. However, it still requires significant memory as it computes probabilities for all inactive parameters. Inspired by this, we propose a more efficient approach: sampling embedding vectors based on their associated user/item frequency and computing sparse gradients exclusively for parameters within this subset. This method ensures that sparsity is maintained for both forward and backward passes. By calculating probabilities only for each user/item and utilizing sparse gradients, SparseRec significantly reduces memory usage while preserving sparsity throughout the training phase. In short, the contribution of our paper can be outlined below:
\begin{itemize}
    \item We introduce a novel NMF-based mask matrix initialization method specially designed for DST operating on GNN-based recommenders.
    \item We empirically demonstrate that the regrowth strategy using instantaneous gradients fails to allocate sufficient parameters to important users/items and propose a novel regrowth strategy to address this issue.
    \item We present a sparse gradient technique that samples a subset of embedding vectors and evaluates gradients only for parameters in this subset.
    \item We combine sparseRec with three base recommenders and conduct empirical evaluations on two datasets. Comparing its performance with various state-of-the-art lightweight embedding algorithms affirm its efficacy for GNN-based recommenders.
\end{itemize}

\section{Related Works}
\subsection{Deep Recommender Systems}
Researchers have proposed various deep recommender models to capture user-item relationships \cite{Yin2024Ondevicerec}. The initial developments in deep recommender systems are based on Matrix Factorization (MF) \cite{He2016eALS, Hu2008ALS, Koren2021SCD++}, which aims to decompose the user-item interaction matrix to derive latent user and item embeddings. He et al. \cite{He2017NCF} advanced this approach by integrating Matrix Factorization with an MLP and proposed NCF. Beyond MF-based techniques, there has been exploration into GNN architectures \cite{Wang2019NGCF, He2020LightGCN, Mao2021UltraGCN}. For example, NGCF \cite{Wang2019NGCF} extends NCF by incorporating graph convolution networks to model user-item interactions. LightGCN \cite{He2020LightGCN} further refines NGCF by removing self-connections, feature transformations, and linear activation functions. LightGCN is further enhanced by UltraGCN \cite{Mao2021UltraGCN} and MGDCF \cite{Hu2022MGDCF}. In this paper, we focus on sparsifying the embedding table of GNN-based models because they outperform MF models such as NCF. Apart from user-item interactions, other data types such as location \cite{Yin2015TRM, Wang2020LLRec, Yin2016SpatioTemp}, sentiment \cite{Wang2017LSARS}, activity level \cite{Zhang2024QAGCF}, text \cite{Li2023Recformer, Zhang2024UOEP}, images \cite{Wang2017VPOI}, videos \cite{Lee2018CDML}, user feedback transitions \cite{Zhang2024Modeltransitions, Wang2024DoNotWait} and social networks \cite{Chen2022Social} have also been utilized.

\subsection{Embedding Pruning}
Embedding pruning remove weights based on some pruning criteria such as weight magnitudes \cite{Janowsky1989Pruning, Strom1997Sparse, Thimm1995Evaluating}, Taylor expansion \cite{Molchanov2019Taylor}, derivatives \cite{LeCun1989Heassian, Mozer1988Relevance, Molchanov2017Pruning}. Liu et al. \cite{Liu2021PEP} proposed to learn the pruning threshold for each parameter in an end-to-end differentiable manner. Despite the diversity of the pruning criteria, they can all be interpreted as learning a salience score for each parameter \cite{Heddes2024AlwaysSparse}. Frankle et al. \cite{Frankle2019Lottery} introduced the lottery ticket hypothesis, stating a sparse sub-network can match the performance of the larger, randomly-initialized dense network when it is trained in isolation. To find such as sub-network, Han et al. \cite{Han2015Lottery} proposed a train-prune-retrain paradigm that trains a dense network, prunes it and retrains it. In an effort to avoid dense training and preserve sparsity throughout training, sparse training methods have been proposed, including static sparse training (SST) \cite{Lee2018Snip, Tanaka2020SynFlow, Wang2020GraSp} and dynamic sparse training (DST) \cite{Evci2020RigL, Dettmers2019Momentum, Wang2024DSL}. SST prunes the dense network at initialization prior to training. DST, on the other hand, iteratively explores weights during training by iteratively prune old weights and activate new ones. SST can be seen as the initialization step in DST. Our proposed method, sparseRec, can be classified as a DST method.

\section{Preliminary}
This section outlines the general framework of DST when it is applied to recommender models as in \cite{Dettmers2019Momentum}. Let $\mathbf{U}$ and $\mathbf{V}$ denote the user and item embedding table, respectively. They can be stacked together to form the overall embedding table $\mathbf{E} = [\mathbf{U}; \mathbf{V}]$. We first initialize the binary mask matrix $\mathbf{M}$, which can be split into a user part $\mathbf{M}^U$ and an item part $\mathbf{M}^V$. We will delve into the step of sparse initialization in Section \ref{sec:init}. The embedding table can thus be sparsified by $\mathbf{E} \circ \mathbf{M}$, where $\circ$ denotes element-wise multiplication. The density ratio $d$ of the embedding table can be expressed as 
\begin{equation}
    d = \frac{\|\mathbf{M}\|_0}{(m + n)s},
\end{equation}
where $m$ is the number of users, $n$ is the number of items, $s$ is the maximal embedding size and $\|\cdot\|_0$ is the number of nonzero elements.
 
Once training begins, DST periodically performs weight exploration that follows a prune-regrow schedule every $\Delta T$ training steps. At the $t$-th step, we first calculate the mean weight magnitude of each user/item embedding table and normalize it with the mean weight magnitude of the entire embedding table:
\begin{equation} \label{eq:mu}
    \mu^U = \frac{\sum_{i=1}^{m} \sum_{j=1}^{s}|\mathbf{U}_{ij}|}{\sum_{i=1}^{m + n} \sum_{j=1}^{s}|\mathbf{E}_{ij}|} \;\;\; \mu^V = \frac{\sum_{i=1}^{m} \sum_{j=1}^{s}|\mathbf{V}_{ij}|}{\sum_{i=1}^{m + n} \sum_{j=1}^{s}|\mathbf{E}_{ij}|}
\end{equation}
where $|\cdot|$ is the element-wise absolute value operation and subscript $ij$ denotes the $(i,j)$-th element of a matrix. Note that $\mu^U$ and $\mu^V$ can be seen as the contribution of each user/item embedding table to the overall mean weight magnitude.

Then we enter the pruning stage at the $t$-th step, when a proportion (with pruning rate $\rho_t$) of active parameters in each user/item embedding table with the smallest magnitudes are selected to be dropped:
\begin{equation} \label{eq:prune}
    \mathcal{P}_t^U = \text{topK}(|\mathbf{U}|, \rho_t\|\mathbf{M}^U\|_0) \;\;\; \mathcal{P}_t^V = \text{topK}(|\mathbf{V}|, \rho_t\|\mathbf{M}^V\|_0),
\end{equation}
where $\mathcal{P}_t^U$ (or $\mathcal{P}_t^V$) denote the set of active parameters to be pruned in the user and item embedding tables at the $t$-th step. Note that $\mathcal{P}^U$ and  $\mathcal{P}^V$ are updated periodically during training, but we omit its subscript $t$ henceforth for simplicity. We also omit this subscript for other variables if there is no ambiguity. After the pruning stage, a total of $\text{card}(\mathcal{P}^U) + \text{card}(\mathcal{P}^V)$ parameters are pruned. $\text{card}(\cdot)$ denotes the cardinality of a set. We can remove elements in $\mathcal{P} = \mathcal{P}^U \cup \mathcal{P}^V$ from $\mathbf{E}$ by setting their corresponding mask in $\mathbf{M}$ to zeros and performing element-wise multiplication.

Cosine annealing is applied to the pruning rate. Let $\rho_0$ denote the initial pruning rate and $T_{end}$ denote the total number of training epochs, the pruning rate at the $t$-th step can be expressed as:
\begin{equation} \label{eq:cosine}
    \rho_t = \frac{\rho_0}{2} \left(  1 + \text{cos}(\frac{\pi t}{T_{end}}) \right)
\end{equation}
 
Following the pruning stage is the regrowth stage, when inactive parameters are reactivated according to a gradient-based regrowth criterion which we will shed light on in Section \ref{sec:sparsegradient} and \ref{sec:cumgrad}. Newly activated parameters are initialized to zeros so they will not affect the output of the base recommender. The number of parameters to be activated in the user/item embedding table is the total number of pruned parameters times their respective magnitude contribution, namely:
\begin{align}
    \text{card}(\mathcal{G}^U) &= \mu^U \Bigl( \text{card}(\mathcal{P}^U)  + \text{card}(\mathcal{P}^V) 
\Bigl) \nonumber \\ 
    \text{card}(\mathcal{G}^V) &= \mu^V \Bigl( \text{card}(\mathcal{P}^U) + \text{card}(\mathcal{P}^V) 
 \Bigl),
\end{align}
where $\mathcal{G}_U$ and $\mathcal{G}_V$ denote the regrowth sets containing the inactive parameters to be activated in the user and item embedding tables.

\begin{figure*}
    \centering
    \includegraphics[width=0.75\textwidth]{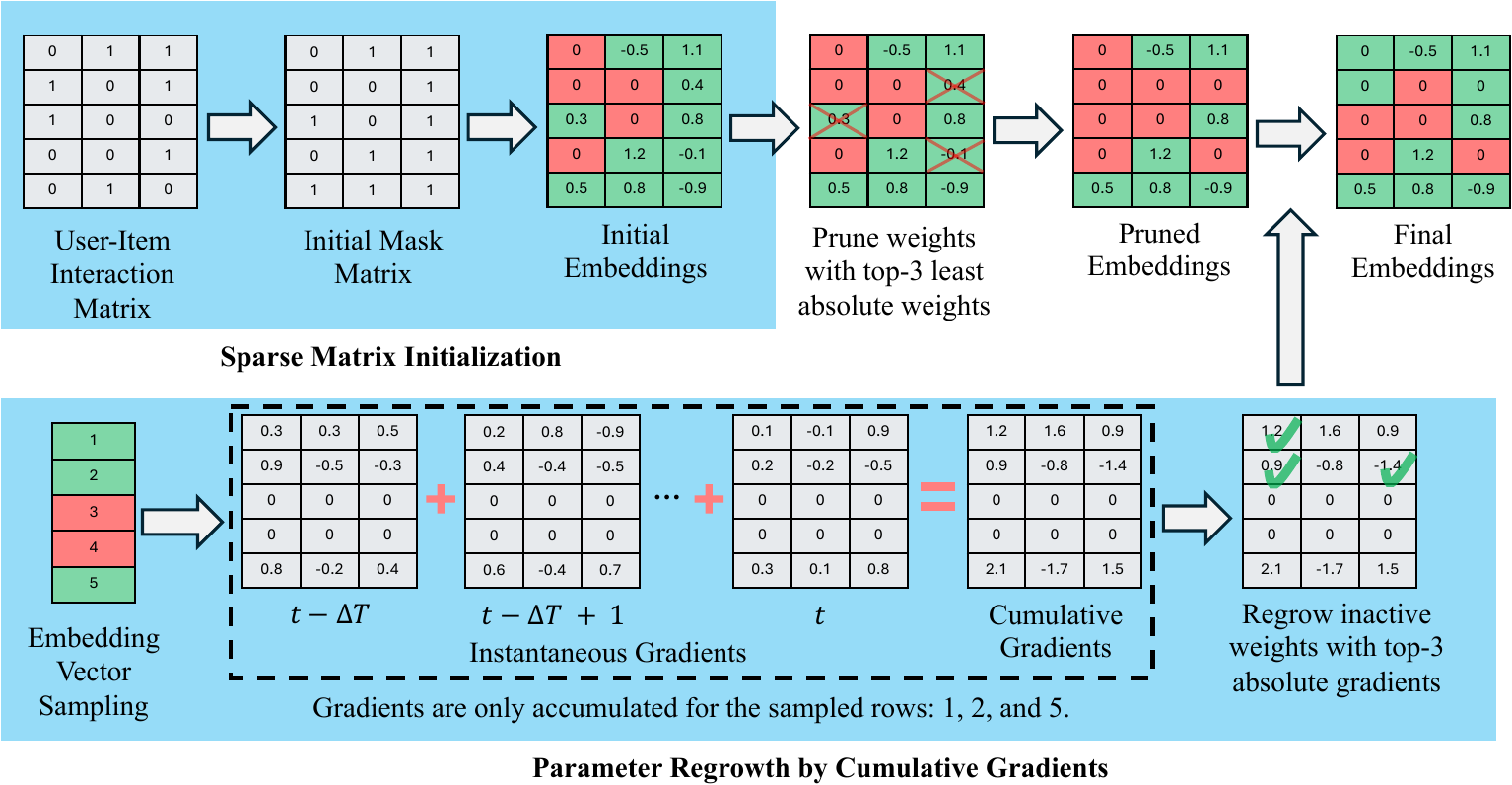}
    \caption{A toy example of sparseRec.}
    \label{fig:overview}
\end{figure*}

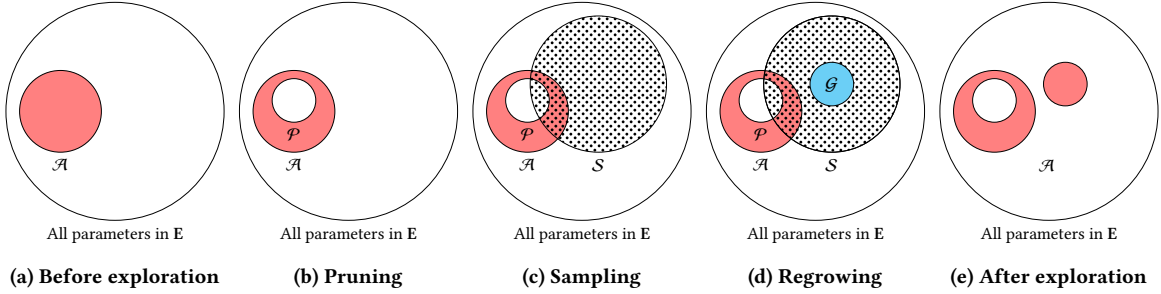
\begin{figure*}[th] 
    \centering
    \begin{subfigure}[b]{0.17\textwidth}
        \centering
        \resizebox{\textwidth}{!}{
            \begin{tikzpicture}
                \node [draw, circle, minimum size=4cm,  label={270:All parameters in $\mathbf{E}$}] [] at (0,0){};
                \node [draw, circle, minimum size=1.5cm, fill=red!50, label={270:$\mathcal{A}$}] (A) at (-1,0){};
            \end{tikzpicture}
        }
        \caption{Before exploration}
    \end{subfigure}
    \begin{subfigure}[b]{0.17\textwidth}
        \centering
        \resizebox{\textwidth}{!}{
            \begin{tikzpicture}
            \node [draw, circle, minimum size=4cm,  label={270:All parameters in $\mathbf{E}$}] [] at (0,0){};
            \node [draw, circle, minimum size=1.5cm, fill=red!50, label={270:$\mathcal{A}$}] (A) at (-1,0){};
            \node [draw, circle, minimum size=0.8cm, fill=white, label={270:$\mathcal{P}$}] (P) at (-1,0.2){};
        \end{tikzpicture}
        }
        \caption{Pruning}
    \end{subfigure}
    \begin{subfigure}[b]{0.17\textwidth}
        \centering
        \resizebox{\textwidth}{!}{
            \begin{tikzpicture}
            \node [draw, circle, minimum size=4cm,  label={270:All parameters in $\mathbf{E}$}] [] at (0,0){};
            \node [draw, circle, minimum size=1.5cm, fill=red!50, label={270:$\mathcal{A}$}] (A) at (-1,0){};
            \node [draw, circle, minimum size=0.8cm, fill=white, label={270:$\mathcal{P}$}] (P) at (-1,0.2){};
            \node [draw, circle, minimum size=2.5cm, pattern=crosshatch dots, label={270:$\mathcal{S}$}] (S) at (0.3,0.5){};
        \end{tikzpicture}
        }
        \caption{Sampling}
    \end{subfigure}
    \begin{subfigure}[b]{0.17\textwidth}
        \centering
        \resizebox{\textwidth}{!}{
            \begin{tikzpicture}
            \node [draw, circle, minimum size=4cm,  label={270:All parameters in $\mathbf{E}$}] [] at (0,0){};
            \node [draw, circle, minimum size=1.5cm, fill=red!50, label={270:$\mathcal{A}$}] (A) at (-1,0){};
            \node [draw, circle, minimum size=0.8cm, fill=white, label={270:$\mathcal{P}$}] (P) at (-1,0.2){};
            \node [draw, circle, minimum size=2.5cm, pattern=crosshatch dots, label={270:$\mathcal{S}$}] (S) at (0.3,0.5){};
            \draw[fill=cyan!50] (0.3,0.5) circle (0.4cm);
            \node[] at (0.3, 0.5){$\mathcal{G}$};
            \draw (0.3, 0.5) circle (1.25cm);
        \end{tikzpicture}
        }
        \caption{Regrowing}
    \end{subfigure}
    \begin{subfigure}[b]{0.17\textwidth}
        \centering
        \resizebox{\textwidth}{!}{
            \begin{tikzpicture}
            \node [draw, circle, minimum size=4cm, label={270:All parameters in $\mathbf{E}$}] [] at (0,0){};
            \draw[fill=red!50] (-1, 0) circle (0.75cm);
            \draw[fill=red!50] (0.3,0.5) circle (0.4cm);
            \node[right] at (-0.3, -1){$\mathcal{A}$};
            \draw[fill=white] (-1, 0.2) circle (0.4cm);
        \end{tikzpicture}
        }
        \caption{After exploration}
    \end{subfigure}
    \caption{Venn diagrams illustrating the relationship between $\mathcal{A}$, $\mathcal{S}$, $\mathcal{G}$ and $\mathcal{P}$ in different stages of weight exploration. $\mathcal{A}$ (red) denotes active parameters; $\mathcal{G}$ (blue) denotes the parameters to be activated, $\mathcal{S}$ (dotted) denotes the sampled parameters, and the white area denotes inactive parameters. $\mathcal{G}$ is mutually exclusive with $\mathcal{A}$ but not necessarily with $\mathcal{P}$.} \label{fig:venn}
\end{figure*}

\section{Methodology}

In a nutshell, our methodology involves three key steps to address the limitations of traditional DST in recommender systems. First, we propose a novel sparse initialization technique that leverages NMF to generate a more effective initial mask matrix, avoiding the pitfalls of random uniform initialization. Second, we employ a cumulative gradient-based regrowth strategy to dynamically allocate parameters to the most impactful users/items, overcoming the shortcomings of the traditional regrowth method that relies on instantaneous gradients. Third, to minimize computational overhead and maintain sparsity throughout training, we introduce a sparse gradient computation mechanism that selectively computes gradients only for a sampled subset of parameters, based on user/item frequency, thus ensuring efficient memory usage while preserving model performance. 

\subsection{Sparse Matrix Initialization} \label{sec:init}

Studies \cite{Peng2022SparseGraph, Evci2020RigL} have shown that random uniform initialization of a sparse network impedes the post-training performance. As discussed in Section \ref{sec:intro}, alternative initialization methods like ER initialization have been proposed for the dense and convolution layers in CNNs and DNNs \cite{Lee2018Snip, Peng2022SparseGraph}, but they are not directly applicable to GNN-based recommender models where the focus is on pruning the embedding table. Gradient-based methods come with additional computational overhead because they entail dense gradients \cite{Lee2018Snip, Wang2020GraSp, Tanaka2020SynFlow}. To tackle this challenge, we propose to use NMF to derive the binary mask $\mathbf{M}$. Matrix Factorization (MF) has been commonly used as recommender systems \cite{He2016eALS, Hu2008ALS, Koren2021SCD++}. Unlike other MF methods, NMF yields naturally sparse representations for users and items. In contrast to random uniform initialization, this data-driven approach can initializes the mask matrix to a local optimum. Concretely, we factorize the user-item interaction matrix $\mathbf{R} \in \mathbb{R}^{m \times n}$ into $\mathbf{W} \in \mathbb{R}^{m \times s}$ and $\mathbf{H} \in \mathbb{R}^{n \times s}$ such that $\mathbf{R} \approx \mathbf{W}\mathbf{H}^T$. By binarizing $\mathbf{W}$ and $\mathbf{H}$, we can obtain $\mathbf{M}^U$ and $\mathbf{M}^V$:
\begin{equation} 
    \mathbf{M}_{ij}^U=
    \begin{cases}
      1 & \text{if}\ \mathbf{W}_{ij} \neq 0 \\
      0 & \text{otherwise}
    \end{cases} \;\;\;
    \mathbf{M}_{ij}^V=
    \begin{cases}
      1 & \text{if}\ \mathbf{H}_{ij} \neq 0 \\
      0 & \text{otherwise}
    \end{cases}
\label{eq:nmf}
\end{equation}
If the resultant density exceeds the desired level, we prune the weights with the smallest magnitudes in $\mathbf{W}$ or $\mathbf{H}$. Conversely, if the density falls short of the target, we enter a regrowth stage after the first epoch and activate enough parameters to reach the target density. Additionally, it is worth noting that the proposed sparse matrix initialization approach only initializes $\mathbf{M}$. The embedding table $\mathbf{E}$ is still initialized randomly. 

\subsection{Sparse Gradients} \label{sec:sparsegradient}

The gradients for inactive parameters are not necessarily zero, so gradients need to be computed for both active and inactive parameters in the regrowth stage. To avoid dense gradients, we can compute gradients for the active parameters plus a subset of inactive parameters that are predicted to be important to performance improvement. Previous studies \cite{Liu2020ESAPN, Liu2021PEP} have shown that frequently occurring features often require larger embedding sizes. Leveraging this insight, we sample a portion (with sampling ratio $\omega$) of embedding vectors without replacement from the user and item embedding tables based on user/item frequencies. This approach allows vectors with higher user/item frequencies to have a greater likelihood of being sampled. The sampling process can be viewed as selecting $\omega m$/$\omega n$ elements from a multinomial probability distribution with probabilities $\mathbf{p}^U$/$\mathbf{p}^U$: 
\begin{equation} \label{eq:sample}
    \mathcal{S}^U = \text{Multinomial}(\mathbf{p}^U, \omega m) \;\;\; \mathcal{S}^V = \text{Multinomial}(\mathbf{p}^V, \omega n),
\end{equation}
where $\mathcal{S}^U$ and $\mathcal{S}^V$ are the sets of embeddings vectors sampled from the user and item embedding tables. The probabilities can be calculated as:
\begin{equation}
    \mathbf{p}^U = \text{Softmax}(\mathbf{f}^U) \;\;\; \mathbf{p}^V = \text{Softmax}(\mathbf{f}^V),
\end{equation}
where $\mathbf{f}^U$ and $\mathbf{f}^U$ are the user and item frequencies. The Softmax function is employed to smooth the probabilities, ensuring that infrequent users/items are also considered. Unlike a deterministic heuristic-based strategy that prioritizes frequent users/items, our stochastic approach promotes a more balanced activation. Let $\mathcal{A}^U$ and $\mathcal{A}^V$ denote the active parameters from the user and item embedding table. Define $\mathcal{A} = \mathcal{A}^U \cup \mathcal{A}^V$ as the set of active parameters in both tables, and $\mathcal{B} = \mathcal{B}^U \cup \mathcal{B}^V$ as the set of inactive parameters sampled from both tables. We only need to compute gradients w.r.t. the parameters in the union of active parameters and the sampled set:
\begin{equation} \label{eq:sparsegrad}
    \nabla_{\mathcal{A} \cup \mathcal{S}} \mathcal{L}_{ij}=
    \begin{cases}
        \nabla \mathcal{L}_{ij} & \text{if}\ ij \in {\mathcal{A} \cup \mathcal{S}} \\
        0 & \text{otherwise}
    \end{cases} 
\end{equation}
where $\nabla_{\mathcal{A} \cup \mathcal{S}} \mathcal{L}$ is the sparse gradients w.r.t. $\mathcal{A} \cup \mathcal{S}$, $\nabla \mathcal{L}$ is the dense gradients and $\mathcal{L}$ is the loss function:
\begin{equation}
    \mathcal{L} = \mathcal{L}_{BPR} + \| \mathbf{E} \circ \mathbf{M} \|^2
\end{equation}

\subsection{Parameter Regrowth by Cumulative Gradients} \label{sec:cumgrad}
Upon entering the regrowth stage at step $t$, traditional DST uses instantaneous gradients at the $t$-th batch to assist parameter activation. Consequently, when they are applied to recommender systems, users/items sampled in the $t$-th batch, which do not necessarily contribute to the model performance a lot, enjoy more parameter activation. To activate parameters for the users/items that truly matter, we accumulate the sparse gradients w.r.t. the parameters in $\mathcal{S}$ from the $(t - \Delta T)$-th to the $t$-th step and store the cumulative gradients in $\mathcal{C}^U$ and  $\mathcal{C}^V$:
\begin{equation}
    \mathbf{C}^U = \sum_{i=t-\Delta T}^{t} \left( \nabla_{\mathcal{S}^U}\mathcal{L} \right)_i \;\;\; 
    \mathbf{C}^V = \sum_{i=t-\Delta T}^{t} \left( \nabla_{\mathcal{S}^V}\mathcal{L} \right)_i
\end{equation}

In the pruning stage, a total number of $\text{card}(\mathcal{P})$ parameters are pruned. Therefore, to maintain target density, we need to regrow the same number of parameters in the regrowth stage. We use the mean magnitude contribution to redistribute parameters across the user and item embedding tables by activating $\mu^U \text{card}(\mathcal{P})$ parameters from the user embedding table and $\mu^V \text{card}(\mathcal{P})$ parameters from the item embedding table with the greatest cumulative gradient magnitudes:
\begin{align} \label{eq:regrowth}
    \mathcal{G}^U &= \text{topk} \Bigl( |\mathbf{C}^U|,\mu^U \text{card}(\mathcal{P})\rho \Bigl) \nonumber \\
    \mathcal{G}^V &= \text{topk}\Bigl ( |\mathbf{C}^V|,\mu^V \text{card}(\mathcal{P})\rho \Bigl)
\end{align}
Upon completion of the regrowth stage, $\mathcal{C}^U$ and $\mathcal{C}^V$ are re-initialized to zero matrices. 

It is important to note that $\mathcal{P}$, $\mathcal{S}$, $\mathcal{A}$, and $\mathcal{G}$ are implemented as binary matrices instead of actual hash sets to conserve memory. As a result, all set operations can be efficiently performed using logical bit-wise operators. The whole weight exploration schedule is depicted by Venn diagrams in Figure \ref{fig:venn} and Figure \ref{fig:overview} provides a toy example of the sparse initialization and weight exploration process. The pseudo code of sparseRec is provided by Algorithm \ref{alg:sparseRec}.

\begin{algorithm}[t]
    \caption{sparseRec}
    \label{alg:sparseRec}
    \begin{algorithmic}[1]
        \State $\mathbf{M}^U, \mathbf{M}^V \leftarrow$ Eq. \ref{eq:nmf} \Comment{NMF sparse initialization}
        \State $\mathbf{U} \leftarrow \mathbf{U} \circ \mathbf{M}^U$ and $\mathbf{V} \leftarrow \mathbf{V} \circ \mathbf{M}^V$; \Comment{Apply masks}
        \For{$t = 1, \cdots, T_{end}$}
            \If{$t == 1$ or $t \text{ mod } \Delta T == 0$ }
                
                \State $\mathcal{S}^U, \mathcal{S}^V \leftarrow$ Eq. \ref{eq:sample} \Comment{Sampling vectors}
                \State $\mathbf{C} \leftarrow \mathbf{0}$; \Comment{Set cumulative sparse gradients to $\mathbf{0}$}
            \EndIf
            \State $\mathbf{C}^U \leftarrow \mathbf{C}^U + \nabla_{\mathcal{S}^U} \mathcal{L}$ and $\mathbf{C}^V \leftarrow \mathbf{C}^V + \nabla_{\mathcal{S}^V} \mathcal{L}$
            \State $\mathbf{U} \leftarrow \mathbf{U} - \eta \nabla_{\mathcal{A}^U \cup \mathcal{S}^U} \mathcal{L}$ and $\mathbf{V} \leftarrow \mathbf{V} - \eta \nabla_{\mathcal{A}^V \cup \mathcal{S}^V} \mathcal{L}$
             \If{$t \text{ mod } \Delta T == 0$}
                \State $\mathcal{A}^U, \mathcal{A}^V \leftarrow \text{getActiveParameters}(\mathbf{M})$
                \State $\mu^U, \mu^V \leftarrow$ Eq. \ref{eq:mu}
                \State $\mathcal{P}^U, \mathcal{P}^V \leftarrow $ Eq. \ref{eq:prune} \Comment{Pruning parameters}
                \State $\mathcal{G}^U, \mathcal{G}^V \leftarrow $ Eq. \ref{eq:regrowth} \Comment{Regrowing parameters}
                
            \EndIf
            \State $\mathbf{M} \leftarrow \text{setActiveParameters}(\mathcal{A} / \mathcal{P} \cup {\mathcal{G}})$
            \State $\mathbf{U} \leftarrow  \mathbf{U} \circ \mathbf{M}^U$ and $\mathbf{V} \leftarrow  \mathbf{V} \circ \mathbf{M}^V$ 
            \State $\rho_t \leftarrow $ Eq. \ref{eq:cosine};
        \EndFor
    \end{algorithmic}
\end{algorithm}

\subsection{Analysis of Memory Complexity} \label{sec:complexity}
Here we provide a theoretical proof that sparseRec has improved memory efficiency compared to other embedding pruning baselines PEP and DSL. The memory usage of sparseRec is primarily dominated by three components: the sparse embedding table $\mathbf{E}$, the sparse gradients $\nabla_{\mathcal{A} \cup \mathcal{S}} \mathcal{L}$ and the cumulative gradient table $\mathbf{C}$. The total number of parameters can be hence expressed as
$\text{params} = ds(m+n) + \text{card}(\mathcal{A} \cup \mathcal{S}) + \text{card}(\mathcal{S})$, which reaches its maximum in the virtually impossible case where $\mathcal{S}$ and $\mathcal{A}$ are mutually exclusive:
\begin{align}
    \text{params} &= ds(m+n) + \text{card}(\mathcal{A} \cup \mathcal{S}) + \text{card}(\mathcal{S}) \\
    &\leq ds(m+n) + \text{card}(\mathcal{A}) + \text{card}(\mathcal{S}) + \text{card}(\mathcal{S}) \nonumber \\ 
    &= (2d + 2\omega)s(m+n)
\end{align}

The memory usage of DSL (traditional DST) is dominated by the sparse embedding table and the dense gradients, resulting in a total of $(d + 1)s(m+n)$ parameters. We can compare it to the memory usage of sparseRec:
\begin{equation}
    (2d + 2\omega)s(m+n) \leq (d + 1)s(m+n) 
    \Rightarrow \omega \leq \frac{1 - d}{2} 
\end{equation}
Therefore, sparseRec uses less memory than conventional DST methods such as DSL when $\omega \leq \frac{1 - d}{2}$.

The memory usage of PEP is dominated by a dense embedding table, the dense gradients and a dense table storing the pruning thresholds, resulting in a total of $3s(m+n)$ parameters, which is higher than sparseRec and DSL. In summary, the space complexity for sparseRec, DSL and PEP are $\mathcal{O}\Bigl ((2d + 2\omega)s(m+n) \Bigl )$, $\mathcal{O} \Bigl((d + 1)s(m+n) \Bigl)$ and $\mathcal{O} \Bigl(3s(m+n) \Bigl)$, respectively.

\section{Experiments}

\begin{table*}[]
\caption{Comparison of Recall@20 (R@20) and NDCG@20 (N@20) scores for various lightweight embedding methods on (a) Gowalla and (b) Yelp, with density ratio set to 0.0625, 0.125 and 0.25. The highest scores are highlighted and the second highest scores are underlined.}
\label{tab:summary}
\resizebox{0.85\textwidth}{!}{%
\begin{tabular}{cccccccccccccc}
\hline
 &  & \multicolumn{6}{c}{(a) Gowalla} & \multicolumn{6}{c}{(b) Yelp} \\ \hline
\multirow{2}{*}{Density} & \multirow{2}{*}{Model} & \multicolumn{2}{c}{XSimGCL} & \multicolumn{2}{c}{MGDCF} & \multicolumn{2}{c}{LightGCN} & \multicolumn{2}{c}{XSimGCL} & \multicolumn{2}{c}{MGDCF} & \multicolumn{2}{c}{LightGCN} \\ \cline{3-14} 
 &  & R@20 & N@20 & R@20 & N@20 & R@20 & N@20 & R@20 & N@20 & R@20 & N@20 & R@20 & N@20 \\ \hline
\multirow{7}{*}{0.0625} & PEP & \textbf{0.1731} & {\ul 0.1335} & {\ul 0.1630} & {\ul 0.1264} & {\ul 0.1539} & {\ul 0.1187} & 0.0886 & 0.0747 & 0.0750 & 0.0629 & 0.0757 & 0.0620 \\
 & BET & 0.1546 & 0.1201 & 0.1506 & 0.1181 & 0.1421 & 0.1113 & 0.0749 & 0.0627 & 0.0751 & 0.0629 & {\ul 0.0763} & {\ul 0.0637} \\
 & CIESS & 0.1529 & 0.1170 & 0.1492 & 0.1164 & 0.1405 & 0.1087 & 0.0736 & 0.0611 & 0.0746 & 0.0625 & 0.0751 & 0.0623 \\
 & DSL & 0.1658 & 0.1275 & 0.1595 & 0.1248 & 0.1404 & 0.1090 & {\ul 0.0896} & {\ul 0.0758} & 0.0766 & {\ul 0.0651} & 0.0693 & 0.0574 \\
 & CERP & 0.1572 & 0.1217 & 0.1560 & 0.1245 & 0.1477 & 0.1146 & 0.0821 & 0.0689 & {\ul 0.0767} & 0.0640 & 0.0760 & 0.0619 \\
 & SD & 0.1174 & 0.0891 & 0.1222 & 0.0954 & 0.1197 & 0.0907 & 0.0690 & 0.0568 & 0.0614 & 0.0511 & 0.0672 & 0.0553 \\ 
 & sparseRec & {\ul 0.1721} & \textbf{0.1344} & \textbf{0.1722} & \textbf{0.1361} & \textbf{0.1579} & \textbf{0.1248} & \textbf{0.0931} & \textbf{0.0791} & \textbf{0.0816} & \textbf{0.0695} & \textbf{0.0849} & \textbf{0.0716} \\ \hline
\multirow{7}{*}{0.125} & PEP & {\ul 0.1818} & {\ul 0.1404} & 0.1838 & 0.1425 & {\ul 0.1680} & {\ul 0.1287} & 0.0925 & 0.0780 & 0.0878 & 0.0735 & 0.0827 & 0.0680 \\
 & BET & 0.1697 & 0.1306 & 0.1794 & 0.1406 & 0.1641 & 0.1273 & 0.0843 & 0.0709 & \textbf{0.0944} & \textbf{0.0802} & {\ul 0.0866} & {\ul 0.0730} \\
 & CIESS & 0.1632 & 0.1260 & 0.1752 & 0.1383 & 0.1637 & 0.1272 & 0.0831 & 0.0698 & 0.0891 & 0.0756 & 0.0842 & 0.0709 \\
 & DSL & 0.1800 & 0.1398 & {\ul 0.1860} & {\ul 0.1467} & 0.1550 & 0.1192 & {\ul 0.0943} & {\ul 0.0800} & 0.0903 & 0.0767 & 0.0736 & 0.0606 \\
 & CERP & 0.1685 & 0.1301 & 0.1798 & 0.1394 & 0.1638 & 0.1263 & 0.0873 & 0.0732 & 0.0874 & 0.0734 & 0.0810 & 0.0665 \\
 & SD & 0.1475 & 0.1139 & 0.1589 & 0.1234 & 0.1518 & 0.1165 & 0.0833 & 0.0699 & 0.0771 & 0.0647 & 0.0774 & 0.0643 \\ 
 & sparseRec & \textbf{0.1822} & \textbf{0.1418} & \textbf{0.1904} & \textbf{0.1497} & \textbf{0.1721} & \textbf{0.1345} & \textbf{0.0976} & \textbf{0.0833} & {\ul 0.0922} & {\ul 0.0789} & \textbf{0.0907} & \textbf{0.0762} \\ \hline
\multirow{7}{*}{0.25} & PEP & 0.1875 & 0.1445 & 0.2005 & 0.1566 & 0.1754 & 0.1347 & 0.0950 & 0.0801 & {\ul 0.1006} & {\ul 0.0842} & 0.0860 & 0.0712 \\
 & BET & 0.1782 & 0.1375 & {\ul 0.2019} & {\ul 0.1590} & {\ul 0.1805} & {\ul 0.1389} & 0.0912 & 0.0770 & 0.0998 & 0.0842 & {\ul 0.0901} & {\ul 0.0762} \\
 & CIESS & 0.1745 & 0.1354 & 0.2000 & 0.1571 & 0.1798 & 0.1383 & 0.0912 & 0.0766 & 0.0985 & 0.0832 & 0.0895 & 0.0753 \\
 & DSL & {\ul 0.1876} & {\ul 0.1456} & 0.2018 & 0.1584 & 0.1663 & 0.1264 & {\ul 0.0986} & {\ul 0.0833} & 0.1002 & 0.0848 & 0.0780 & 0.0634 \\
 & CERP & 0.1755 & 0.1359 & 0.1997 & 0.1553 & 0.1741 & 0.1341 & 0.0918 & 0.0770 & 0.0996 & 0.0835 & 0.0864 & 0.0712 \\
 & SD & 0.1733 & 0.1341 & 0.1957 & 0.1514 & 0.1717 & 0.1325 & 0.0899 & 0.0752 & 0.0983 & 0.0829 & 0.0819 & 0.0680 \\ 
 & sparseRec & \textbf{0.1879} & \textbf{0.1470} & \textbf{0.2079} & \textbf{0.1634} & \textbf{0.1823} & \textbf{0.1424} & \textbf{0.1003} & \textbf{0.0854} & \textbf{0.1034} & \textbf{0.0884} & \textbf{0.0949} & \textbf{0.0797} \\ \hline
 & DHE & 0.1329 & 0.1016 & 0.2039 & 0.1592 & 0.1450 & 0.1112 & 0.0725 & 0.0450 & 0.1022 & 0.0869 & 0.0734 & 0.0606 \\ \hline
\end{tabular}%
}
\end{table*}

In this section, we aim to answer the following research questions:
\begin{itemize}
    \item \textbf{RQ1}: How does sparseRec perform compared to other state-of-the-art lightweight embedding methods?
    \item \textbf{RQ2}: Can NMF-based sparse initialization improve the performance of sparseRec?
    \item \textbf{RQ3}: Can parameter regrowth by cumulative gradients improve the performance of sparseRec?
    \item \textbf{RQ4}: Is there any correlation between the user/item frequency and embedding size?
\end{itemize}

\subsection{Experimental Setup}
\subsubsection{Datasets} Each lightweight embedding method is evaluated on the user rating data in two datasets: Gowalla and Yelp2018 as in \cite{He2020LightGCN, Wang2019NGCF, Hu2022MGDCF}, both of which are commonly used by GNN-based recommenders. The Gowalla dataset comprises 1,027,370 interactions involving 29,858 users and 40,981 items and the Yelp dataset contains 1,561,406 interactions among 31,668 users and 38,048 items. Both datasets are split into training, validation, and test sets with a ratio of 70\%, 10\%, and 20\%, respectively. We treat all observed user-item interactions as positive and unobserved interactions as negative.

\subsubsection{Baselines} We compare sparseRec against several approaches, including embedding pruning methods (PEP \cite{Liu2021PEP} and DSL \cite{Wang2024DSL}), variable-size embedding methods (CIESS \cite{Qu2023CIESS} and BET \cite{Qu2024BET}), parameter sharing methods (CERP \cite{Liang2023CERP} and DHE \cite{Kang2021DHE}) and conventional small dense networks. Since DHE does not use an embedding table, we are unable to fix its density. Therefore, we search $\{8, 16, 32, 64, 128\}$ for its best output embedding size and present its best performance.

\subsubsection{Base Recommenders} To demonstrate its effectiveness with various state-of-the-art GNN recommenders, we integrate SparseRec with XSimGCL \cite{Yu2024XSimGCL}, MGDCF \cite{Hu2022MGDCF} and LightGCN \cite{He2020LightGCN}. We apply as the base recommenders. Each method with adjustable embedding density ratio is tested with the mean embedding size set to 8, 16 and 32, corresponding to density ratios of 6.25\%, 12.5\% and 25\% when the full embedding size is 128. The combination of three base recommenders, two datasets and three density ratios lead to 18 experimental settings.

\subsubsection{Metrics} We evaluate the performance of each method using Recall@20 and NDCG@20, which are common evaluation metrics in recommendation \cite{He2020LightGCN, Wang2019NGCF, Hu2022MGDCF}.

\subsection{Implementation Details}
Here we provide the implementation details of sparseRec. Early stopping is implemented after 300 training epochs, evaluating model performance on the validation set every 5 epochs, with a patience threshold of 5 epochs. The total number of training epochs is $T_{end}$ is 500. NMF is solved using the Coordinate Descent solver. Learning rate decay is used with the initial learning rate set to 0.01. It decays after every epoch by a factor of 0.99 when the LightGCN base recommender is train on the Yelp dataset. The decay rate is  0.995 in all other settings. The minimal learning rate is 0.0005. The full embedding size $s$ is 128. Bayesian Personalized Ranking Loss \cite{Rendle2009BPR} is used to train the recommenders. Adam optimizer \cite{Kingma2014Adam} is used with weight decay of 0.0001. The batch size is 8000. For the base recommenders, we inherit their optimal settings as reported in their original papers. After training, the embedding table is quantized into signed 8-bit integers as quantization is a common practice in the real world. During testing, it is converted back to floating-point format.

\subsection{Overall Performance Comparison}

Table \ref{tab:summary} presents the Recall@20 and NDCG@20 scores for various lightweight embedding methods across different base recommender models and density ratios. Our proposed method achieves state-of-the-art performance in 16 out of the 18 settings, effectively addressing RQ1. The largest performance gain over the second-best baseline is 11.79\%, achieved on the Yelp dataset with a 0.0625 density ratio and the LightGCN base recommender. Other methods such as DSL, PEP and BET also show competitive performance in many settings. For the settings where sparseRec achieves the best performance, we perform paired student's $t$-test between sparseRec and the best baseline method to further affirm the advantageous efficacy of sparseRec. With all $p$-values being less than 0.02, we can confidently conclude that the superior performance of sparseRec is statistically significant and not due to random chance.

In most cases, we observe that the performance improvement of sparseRec over SD is more substantial in low-density settings, where efficient parameter utilization is crucial to minimizing performance degradation. This highlights sparseRec's robustness in memory-constrained scenarios such as mobile and WoT devices.

Moreover, although DHE employs fewer parameters than all other methods by eliminating the conventional embedding table, it exhibits the worst performance in top-$k$ recommendation.

\subsection{Model Component Analysis}
Here we showcase the performance efficacy of the core components of sparseRec, which are our main contributions. MGDCF is chosen as the base recommender as it has the best performance. The density ratio is set to 0.125.

\begin{table}
\caption{Performance of different variations of sparseRec.}
\label{tab:ablation}
\resizebox{\columnwidth}{!}{%
\begin{tabular}{cccccc}
\hline
 &  & \multicolumn{2}{c}{(a) Gowalla} & \multicolumn{2}{c}{(b) Yelp} \\ \hline
regrowth & Initialization & R@20 & N@20 & R@20 & N@20 \\ \hline
Cumulative & NMF & \textbf{0.1904} & \textbf{0.1497} &\textbf{0.0922} & \textbf{0.0789} \\ 
Instantaneous & NMF & 0.1857  & 0.1462 & 0.0902 & 0.0770 \\
Cumulative & Uniform & 0.1809 & 0.1433 & 0.0885 & 0.0754 \\
Instantaneous & Uniform & 0.1754 & 0.1386 & 0.0842 & 0.0717 \\ 
\hline
\end{tabular}
}
\end{table}

\subsubsection{NMF-based Sparse Initialization}
To investigate the influence of NMF sparse initialization, we run two versions of sparseRec: one initializes the sparse network using random uniform distribution as used in traditional DST and the other does so using NMF. Table \ref{tab:ablation} shows NMF initialization improves the performance of sparseRec, thus answering RQ2. 

\subsubsection{Regrowth by Cumulative Gradient}

\begin{figure}[h]
    \begin{subfigure}{\columnwidth}
        \includegraphics[width=0.85\columnwidth]{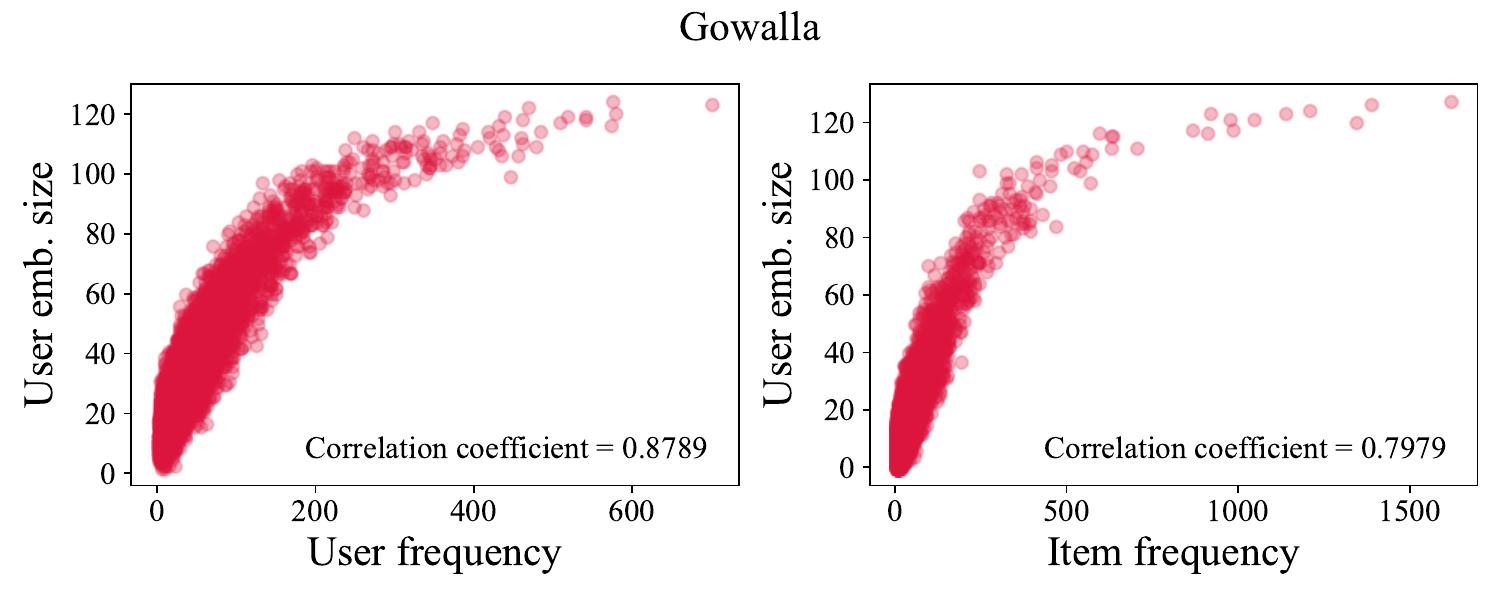}
 
        \includegraphics[width=0.85\columnwidth]{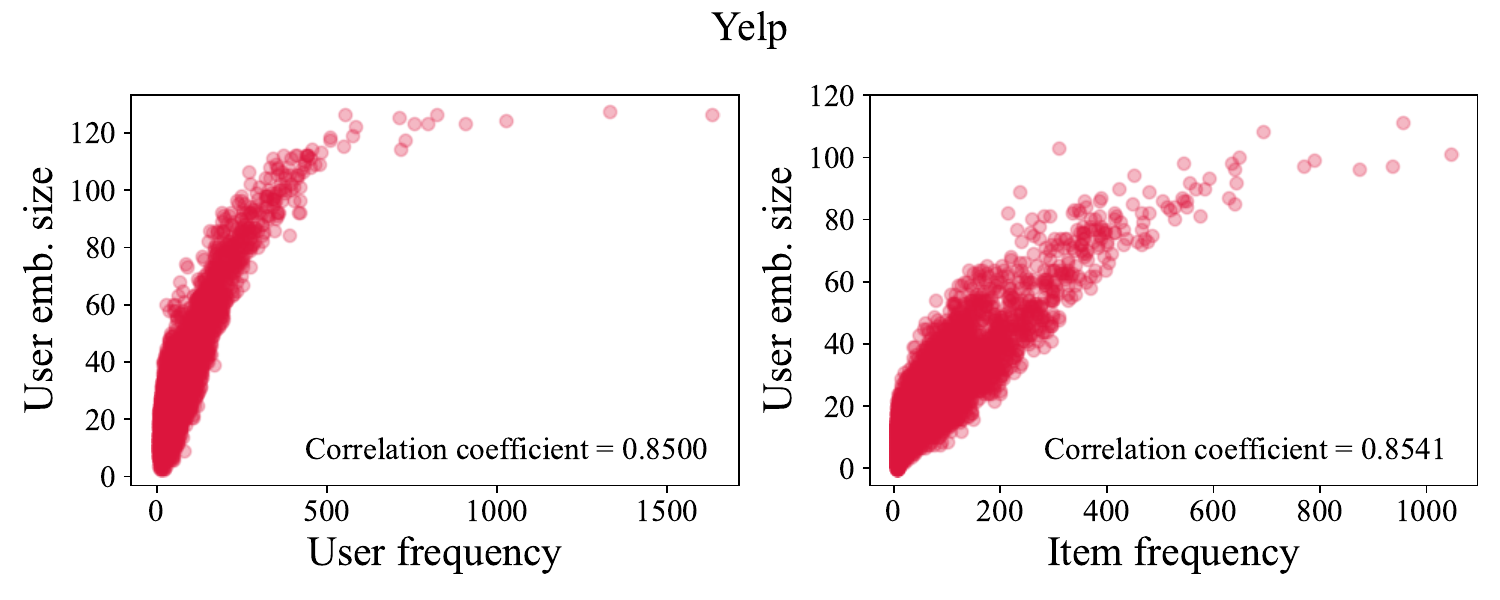}
        \vspace{-1em}
        \caption{regrowth by cumulative gradients}
    \end{subfigure}
    \begin{subfigure}{\columnwidth}
        \includegraphics[width=0.85\columnwidth]{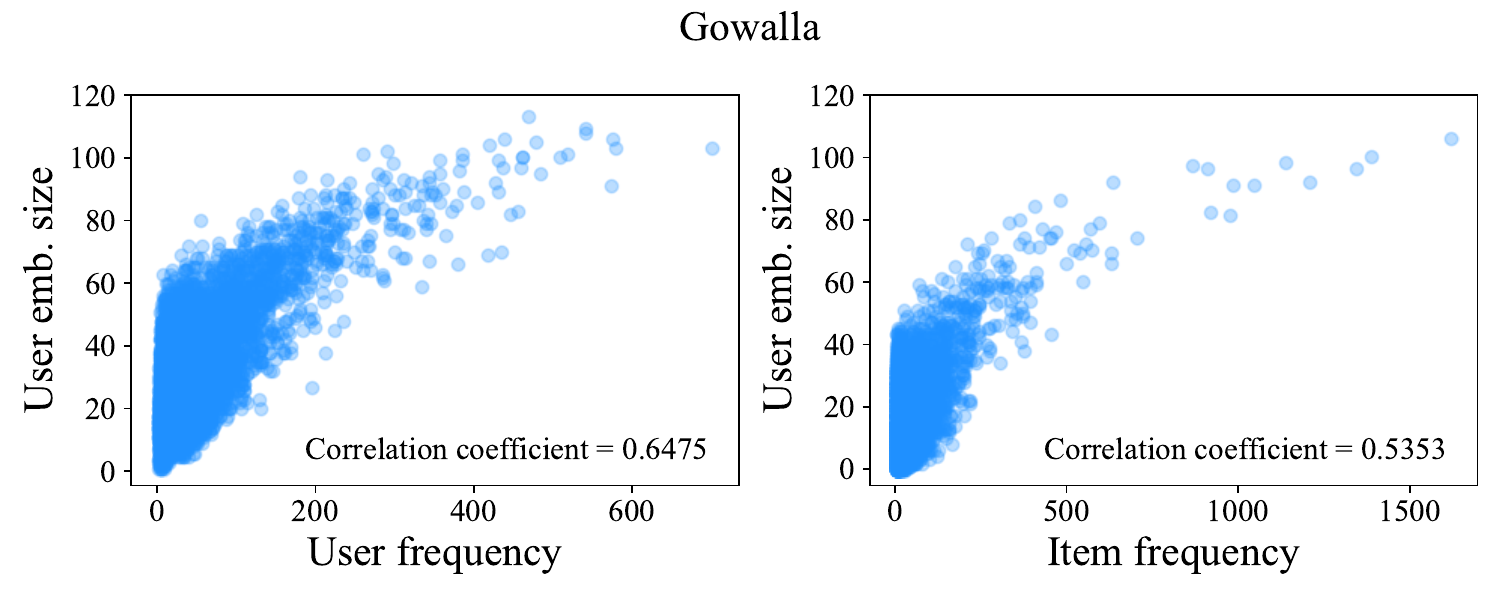}
        \includegraphics[width=0.85\columnwidth]{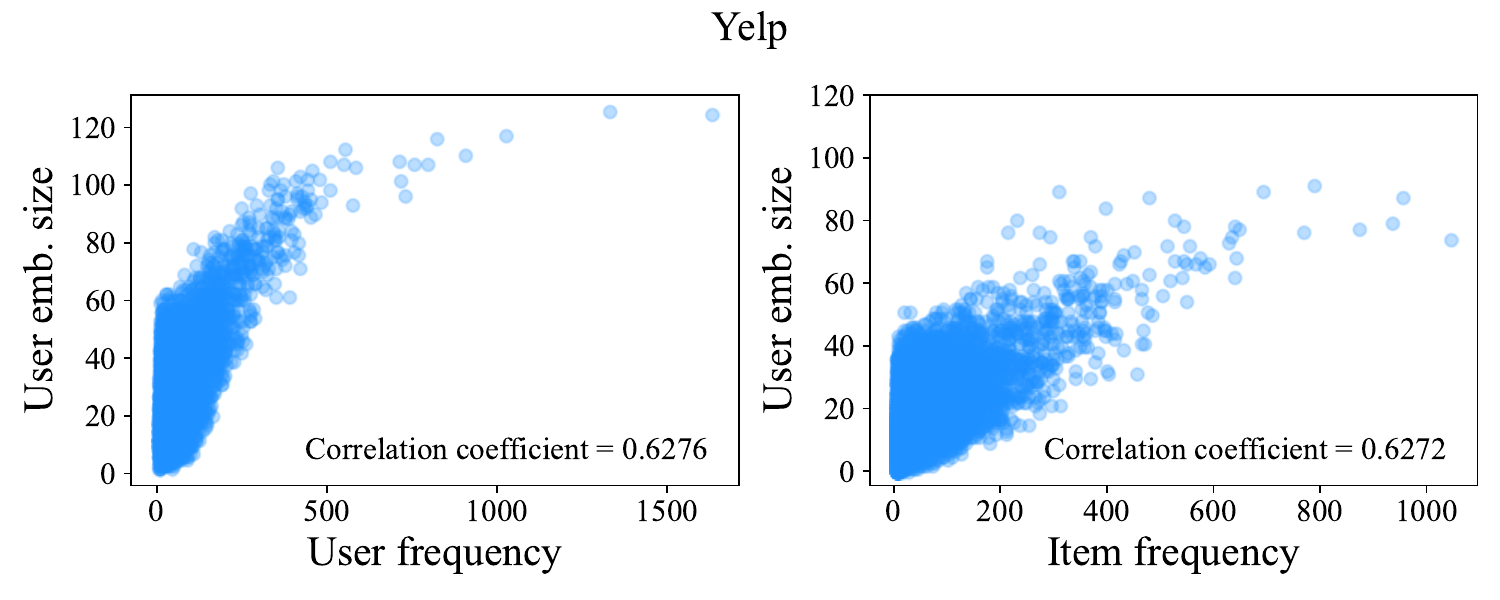}
        \vspace{-1em}
        \caption{regrowth by instantaneous gradients}
    \end{subfigure}
    \caption{The relationship between frequency and embedding sizes in regrowth by (a) cumulative gradients and (b) instantaneous gradients with their respective Pearson Correlation Coefficient. }
    \label{fig:cumgrad}
    \vspace{-1em}
\end{figure}

To illustrate the impact of regrowing parameters based on cumulative gradients versus instantaneous gradients, we run two versions of \texttt{sparseRec}: one using cumulative gradients and the other using instantaneous gradients as used in traditional DST. As shown in Table \ref{tab:ablation}, the version using cumulative gradients results in better performance, thus answering RQ3.

To fully understand this performance boost, we calculate and plot the embedding size for each user and item against their frequency. We also calculate the Pearson Correlation Coefficient between the embedding sizes and frequency scores in each case to quantify the strength of correlation. To exclude the impact of the frequency-based sampling, we set the sampling ratio to 1 and use a dense cumulative gradient table. We present result in Figure \ref{fig:cumgrad}. We can see both regrowth strategies result in a positive correlation between frequency and embedding size, thus answering RQ4. This finding is consistent with the findings in \cite{Liu2020ESAPN}. However, the frequency-size correlation is weaker when instantaneous gradients are used, resulting in a larger number of frequent users/items being assigned smaller embedding sizes. This suggests that parameter regrowth based on instantaneous gradients leads to less reliable regrowth and fails to properly identify important parameters.

\begin{figure}[th]
    \begin{subfigure}{0.85\columnwidth}
        \includegraphics[width=\columnwidth]{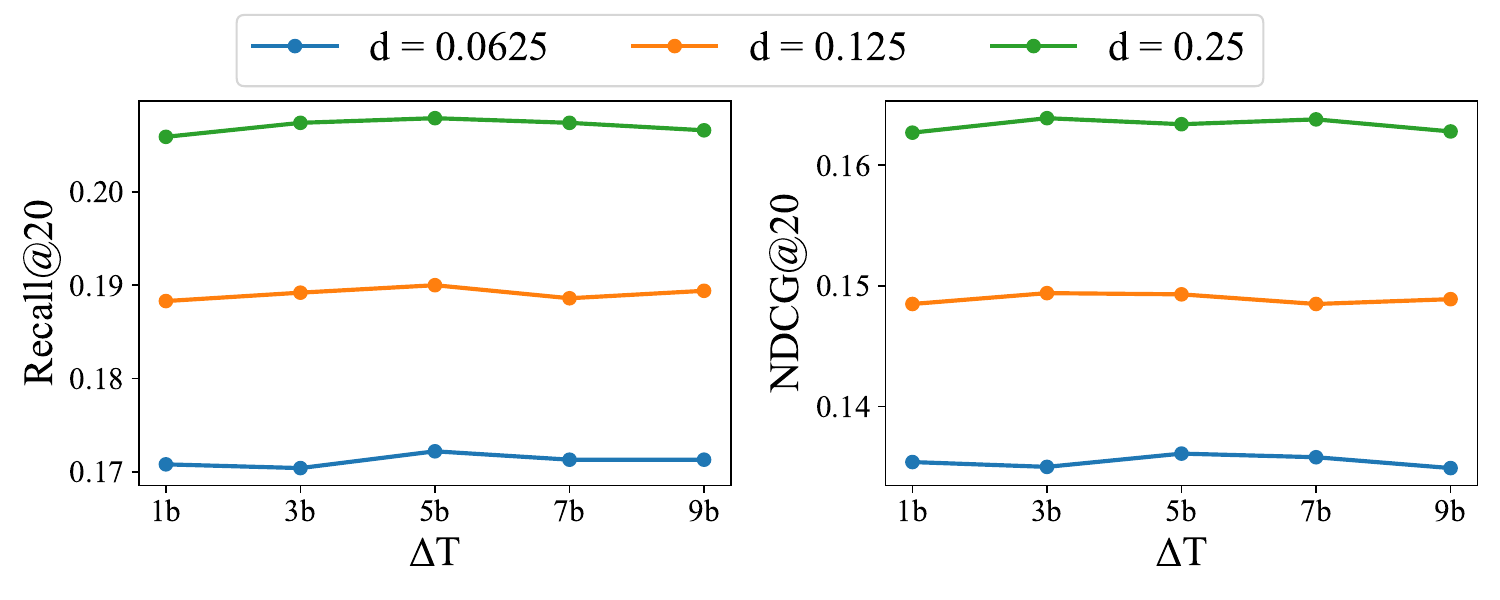}
        \vspace{-2em}
        \caption{Gowalla}
    \end{subfigure}
    \begin{subfigure}{0.85\columnwidth}
        \includegraphics[width=\columnwidth]{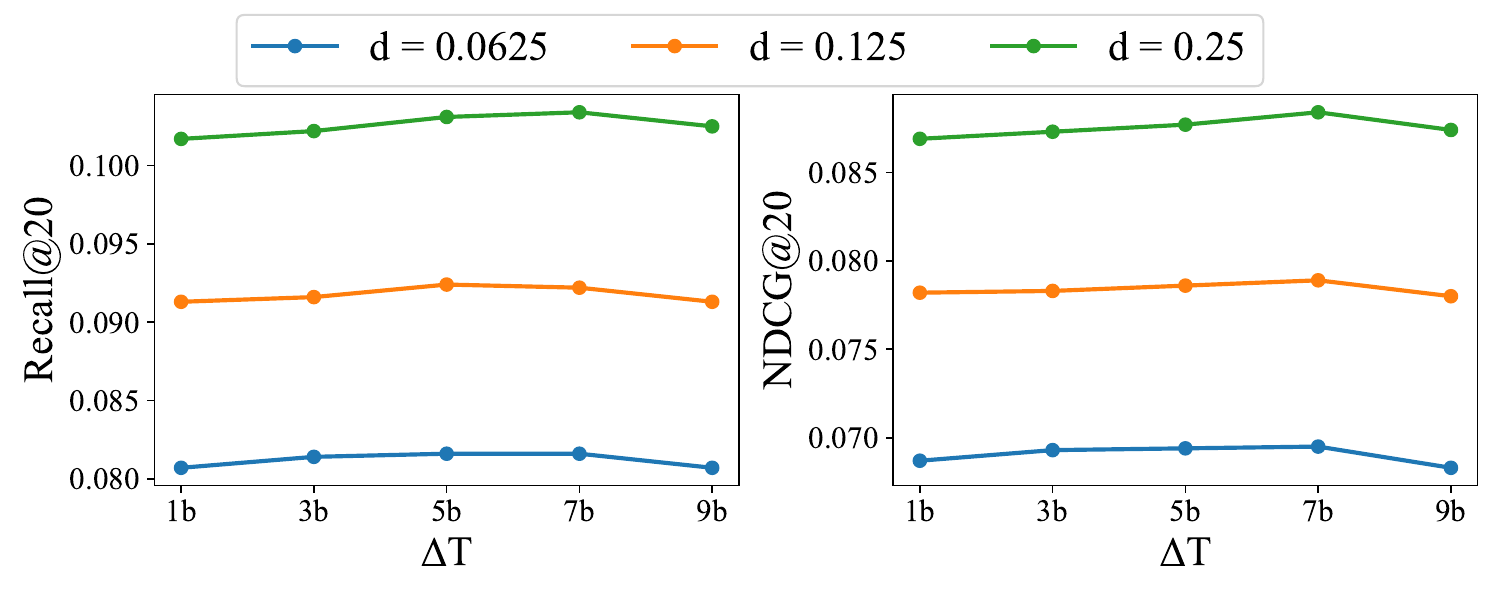}
        \vspace{-2em}
        \caption{Yelp}
    \end{subfigure}
    \caption{Hyerparameter analysis of the weight exploration interval $\Delta T$ w.r.t. Recall@20 and NDCG@20 on (a) Gowalla and (b) Yelp.}
    \label{fig:pruninginterval}
\end{figure}

\subsection{Analysis of Hyperparameter}
In this section, we analyze the influence of key hyperparameters. MGDCF is chosen as the base recommender as it has the best performance. 

\subsubsection{Weight Exploration Interval $\Delta T$}

The weight exploration interval specifies the number of training steps between every weight exploration process and it has a vital influence on the performance of DST. If $\Delta T$ is too small, the parameters are not updated enough to exceed the pruning threshold; if $\Delta T$ is too large, the parameters are not adequately explored and will result in inferior performance \cite{Liu2021IOP}. When $\Delta T -> \infty$, it is equivalent to static sparse training, where we train a base recommender that has been pruned at initialization \cite{Wang2024DSL}. Let $b$ be the number of steps in an epoch. To identify the optimal value for $\Delta T$, we it to $1b$, $3b$, $5b$, $7b$ and $9b$. Then we plot the resulting post-training performance in terms of Recall@20 and NDCG@20 scores under all density ratios $d \in {0.0625, 0.125, 0.25}$. Figure \ref{fig:pruninginterval} shows that sparseRec performs best in the majority of settings when $\Delta T$ is set to $7b$ on the Yelp dataset and $5b$ on the Gowalla dataset. 

\subsubsection{Sampling Ratio $\omega$}
\begin{figure}[h]
    \begin{subfigure}{\columnwidth}
        \includegraphics[width=0.85\columnwidth]{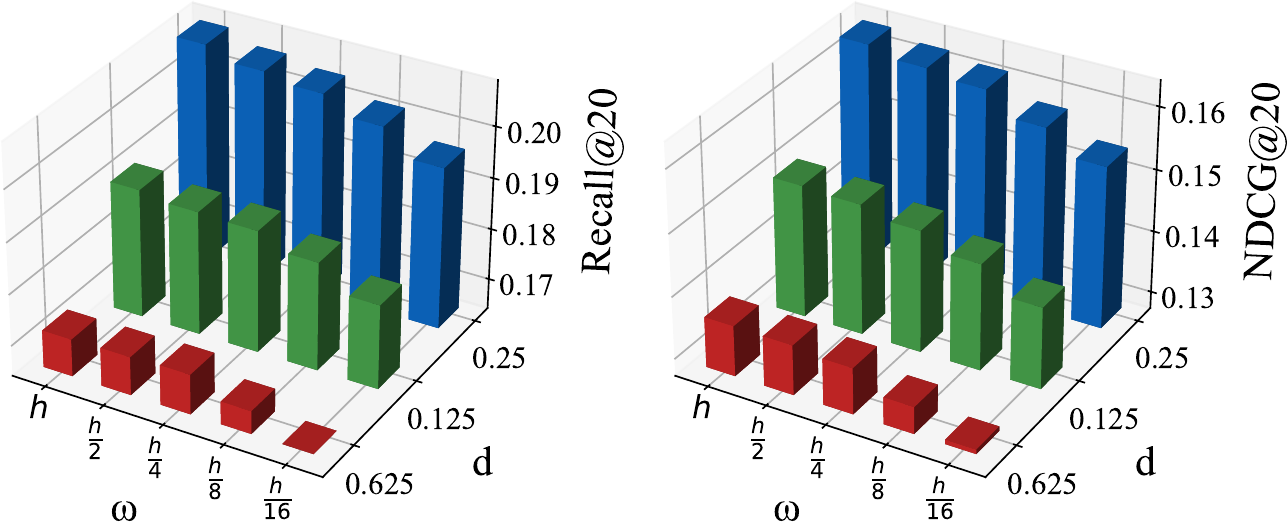}
        \caption{Gowalla}
    \end{subfigure}
    \begin{subfigure}{\columnwidth}
        \includegraphics[width=0.85\columnwidth]{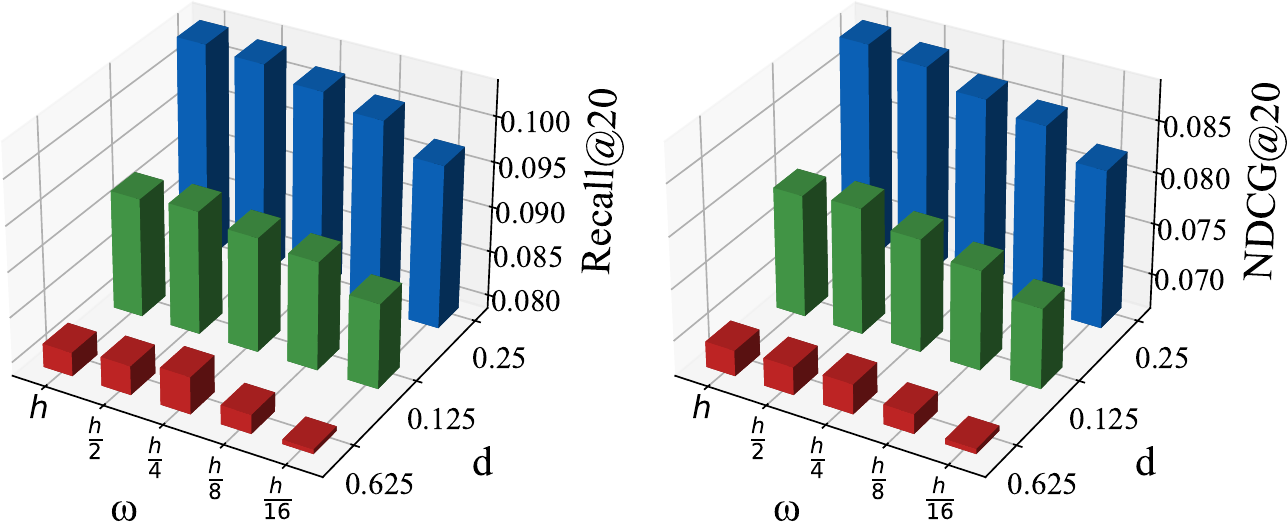}
        \caption{Yelp}
    \end{subfigure}
    \caption{Hyerparameter analysis of the sampling ratio $\omega$ w.r.t. Recall@20 and NDCG@20 on (a) Gowalla and (b) Yelp.}
    \label{fig:omega}
\end{figure}

\begin{figure}[h]
    \includegraphics[width=0.85\columnwidth]{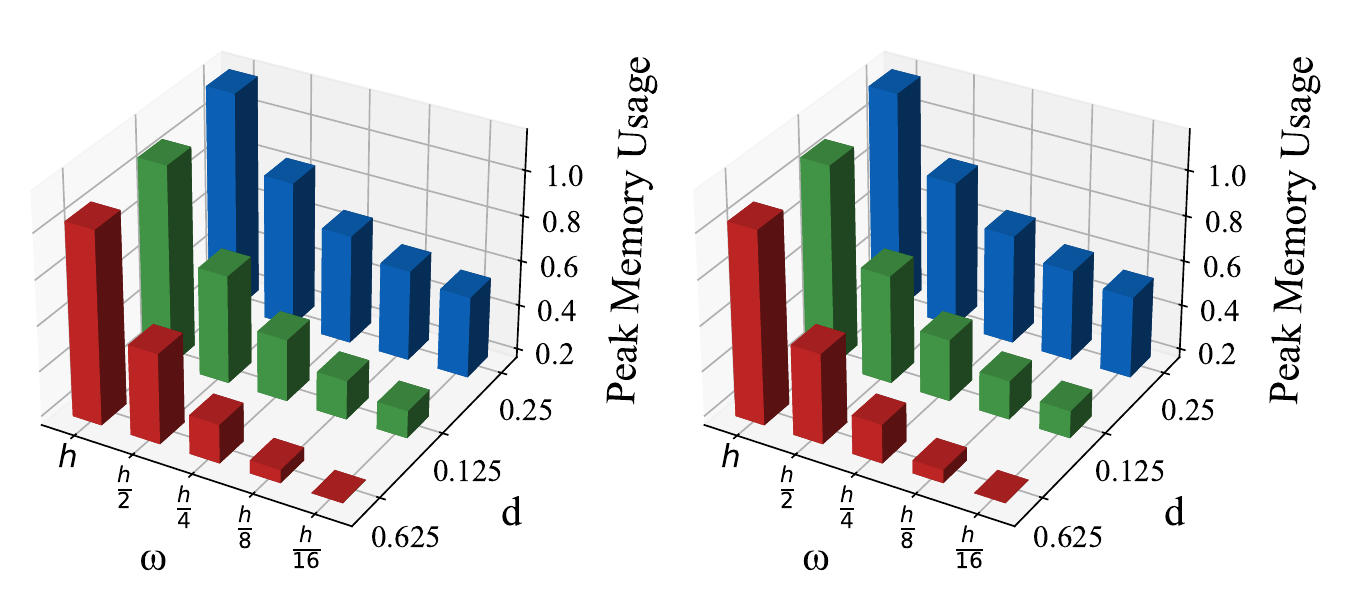}
    \makebox[\columnwidth][c]{%
        \makebox[0.5\columnwidth][c]{\textbf{\small (a) Gowalla}}%
        \makebox[0.5\columnwidth][c]{\textbf{\small (b) Yelp}}%
    }
    \caption{Hyerparameter analysis of the sampling ratio $\omega$ w.r.t. the peak memory usage on (a) Gowalla and (b) Yelp.}
    \label{fig:memory_usage}
\end{figure}

The sampling ratio $\omega$ plays a crucial role in determining the proportion of embedding vectors selected from the embedding table, thereby limiting gradient calculations to only those sampled vectors. As outlined in Section \ref{sec:complexity}, $\omega$ should not exceed $\frac{1-d}{2}$ to ensure that memory savings are realized compared to traditional DST. If $\omega$ is too high, the memory savings diminish, whereas an $\omega$ close to zero essentially results in random regrowth. We define $h = \frac{1-d}{2}$ and plot the recommendation performance metrics Recall@20 and NDCG@20 against peak memory usage in Figure \ref{fig:omega}. The peak memory usage is normalized by the size of a dense embedding table, i.e., $\frac{ds(m+n) + \text{card}(\mathcal{A} \cup \mathcal{B}) + \text{card}(\mathcal{S})}{s(m+n)}$. The analysis is conducted for $\omega$ values in $\{h, \frac{h}{2}, \frac{h}{4}, \frac{h}{8}, \frac{h}{16}\}$ across different density ratios $d \in \{0.25, 0.125, 0.0625\}$. Each plotted point is annotated with its respective sampling ratio. Meanwhile, we also plot the relationship between peak memory usage against different density ratios $d$ and sampling ratios $\omega$ in Figure \ref{fig:memory_usage}. 

As illustrated in Figure \ref{fig:omega}, SparseRec demonstrates improved recommendation performance with larger $\omega$ in most cases, which naturally results in increased memory consumption. A notable observation is that the recommendation performance remains relatively stable even as the sampling ratio is halved from $h$ to $\frac{h}{2}$, particularly with low density ratios. The performance with a mean embedding size of 8 on the Yelp dataset even slightly improves as the sampling ratio decreases from $h$ to $\frac{h}{4}$. This phenomenon might be attributed to the fact that a lower sampling ratio effectively blocks parameters associated with infrequent users/items in the regrowth stage. However, a significant drop in recommendation performance is observed across all density ratios when $\omega$ falls below $\frac{h}{4}$. From Figure \ref{fig:memory_usage}, we also observe that higher density ratios and sampling ratios result in higher peak memory usage.

\section{Conclusion and Future Research}

Dense embedding tables limit the memory efficiency and scalability of recommender systems in mobile and WoT environments with constrained computational resources. To address over-parameterization and reduce memory usage, we introduce sparseRec, which enhances embedding pruning methods by overcoming DST limitations. sparseRec uses NMF for sparse embedding table initialization, calculates sparse gradients for active and sampled parameters during forward and backward passes, and employs cumulative gradients for selective parameter regrowth. Experiments on two real-world datasets with three base recommenders demonstrate sparseRec’s effectiveness, making it ideal for mobile and WoT applications. Future work will explore lightweight Large Language Model-based recommenders for resource-constrained environments.

\clearpage 
\bibliographystyle{ACM-Reference-Format}
\bibliography{sparseRec}

\end{document}